\documentclass[12pt]{iopart}

\usepackage{amsfonts,cancel,xcolor,graphicx,subfigure,url}

\begin{document}

\title{Distributed mean curvature on a discrete manifold for Regge calculus}

\author{Rory Conboye, Warner A Miller, Shannon Ray}

\address{Department of Physics, Florida Atlantic University, Boca Raton,
FL 33431-0991}

\begin{abstract}
The integrated mean curvature of a simplicial manifold is well understood in both Regge Calculus and Discrete Differential Geometry. However, a well motivated pointwise definition of curvature requires a careful choice of volume over which to uniformly distribute the local integrated curvature. We show that hybrid cells formed using both the simplicial lattice and its circumcentric dual emerge as a remarkably natural structure for the distribution of this local integrated curvature. These hybrid cells form a complete tessellation of the simplicial manifold, contain a geometric orthonormal basis, and are also shown to give a pointwise mean curvature with a natural interpretation as a fractional rate of change of the normal vector.

\

\end{abstract}




\section{Introduction}

In representing the curvature of a manifold in some ambient space, the extrinsic curvature gives a measure for the deviation of normal vectors to the manifold about a given point. This deviation can be measured in a number of ways, including: $(i)$ the difference between a parallel transported normal from the normal at a nearby point, $(ii)$ the fractional rate of change of the area of a small region along a one-parameter family of manifolds orthogonal to the normal vectors, or $(iii)$ the Lie derivative of the intrinsic metric along the normal, for the same one-parameter family. Each of these three definitions yields the same extrinsic curvature tensor on taking the infinitesimal limit for a smooth manifold.

In both Regge Calculus (RC) and Discrete Differential Geometry (DDG), smooth manifolds are approximated by piecewise-linear simplicial manifolds. These are formed by connecting blocks of flat space, with each block given by an $n$-simplex. For each of these, the interior geometry is Euclidean, or Minkowski for $4$-simplices in RC, which was developed specifically to give a discrete version of General Relativity (GR). These $n$-simplices are then connected by identifying their $(n-1)$-dimensional faces.

Unlike the continuum case, the different ways of defining curvature on a \emph{piecewise-flat} manifold do not necessarily agree. However, the integral of both the intrinsic and extrinsic curvatures, over certain paths, tend to be well-defined and consistent with the continuum. In the original work of Regge  \cite{Regge}, it was shown that the integrated Gaussian curvature around co-dimension $2$ simplices is all that is required to give an action corresponding with the Einstein-Hilbert action of continuum GR. Any volume that is used to distribute the integrated curvature over, to give a field tensor, cancels and so the action is invariant to such a choice. A similar integrated mean \emph{extrinsic} curvature was first found by Hartle \& Sorkin \cite{HS81}, in extending the Regge action to include boundary terms. These integrated curvatures have also been studied in the DDG community, see \cite{DDGSul} and references therein, with the expression for the integrated mean curvature appearing as far back as 1840 in the work of Steiner \cite{Steiner}.

Though the integrated curvatures are all that is required for defining the action in a vacuum spacetime, a pointwise curvature is necessary for the inclusion of matter. Such a definition requires the choice of some set of volumes to distribute the local integrated curvature over. This leaves room for ambiguities to arise. When the simplicial manifold forms a Delaunay triangulation, it's dual Voronoi lattice is formed by joining the circumcenters of $n$-simplices. This circumcentric dual has long been known as providing one of the most natural dual areas for the intrinsic curvature \cite{M86bbp,ChristFriedLee,HambWill,KLM89ns}. More recent work has used hybrid volumes, depending on both the Delaunay and Voronoi lattices, to define pointwise scalar curvature and pointwise Riemann and Ricci curvature tensors \cite{MMR,SRT}. This has even led to a discrete version of the Ricci flow \cite{SRF}.

Here it is shown that a particular type of hybrid volume, constructed using both the simplicial lattice and its circumcentric dual lattice, gives a remarkably rich interpretation for a pointwise extrinsic curvature. The geometry of these volumes is completely determined by the edge-lengths of the simplicial manifold, they give a complete tessellation of the manifold and each contains a natural geometric orthonormal frame. The distributed mean curvature can also be viewed as a fractional change in normal over a specific displacement.

In what follows, the theory is first developed for the most simple case of a $2$-dimensional surface in $\mathbb{E}^3$. The details of a simplicial submanifold are outlined, and an expression for the extrinsic curvature associated with each edge is derived in section \ref{sec:2D}. The circumcentric dual and hybrid volumes are then introduced in section \ref{sec:dual}, along with the new distributed curvature and its fractional interpretation. More general extensions are considered in section \ref{sec:HigherD}, with the intrinsic properties of simplicial manifolds of higher dimension described in section \ref{sec:HigherDS}, defining the circumcentric dual and hybrid volumes. Hypermanifolds of $\mathbb{E}^n$ are specifically treated in section \ref{sec:hyper}, with the interpretations from the $2$-dimensional case requiring little alteration. Generalizations to arbitrary dimensional submanifolds, and submanifolds of non-Euclidean manifolds are then brief outlined in section \ref{sec:HDgeneral}. 

\section{Simplicial Surfaces in 3-Dimensional Euclidean Space}
\label{sec:2D}

A piecewise-flat simplicial surface $S$, embedded in a 3-dimensional Euclidean space $M$, is a surface which can be decomposed into a collection of flat triangles (2-simplices). The flat geometry of each triangle is completely determined by the length of its edges $l$, which are given by the Euclidean 3-metric. This allows the internal geometry of the surface to be completely determined by the lengths of all of the edges.

\begin{figure}[h]
\begin{center}
\subfigure[Simplicial Surface]{
\includegraphics[scale=0.5]{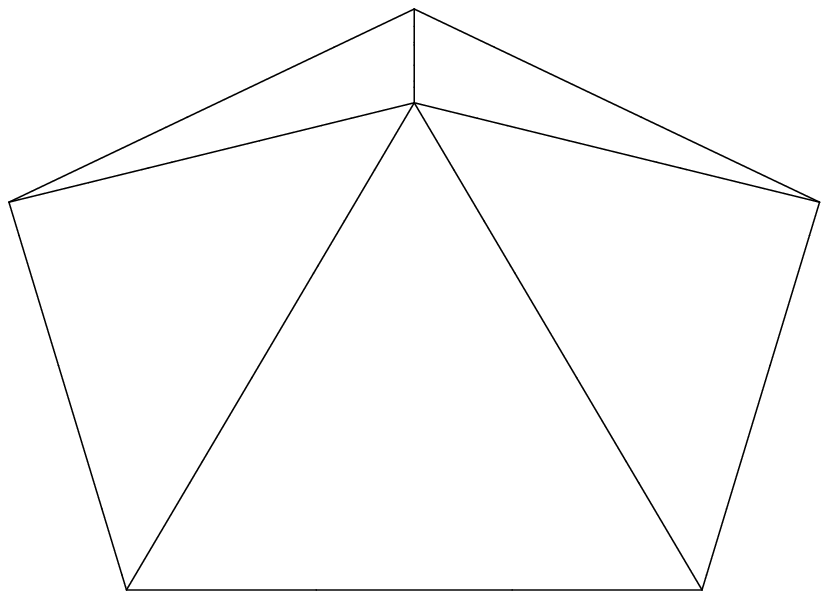} 
\label{fig:SimpSurf}
}
\subfigure[Deficit Angle]{
\includegraphics[scale=0.5]{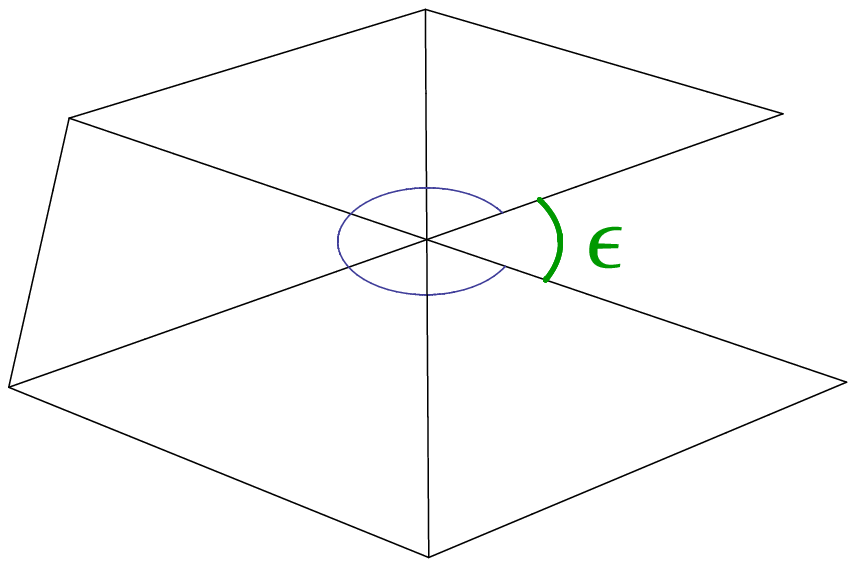} 
\label{fig:DeficitAng}
}
\end{center}
\caption{A simplicial surface consisting of 5 equilateral triangles (a), and the development of one of the triangles (b), giving the deficit angle $\epsilon$ around the central vertex.}
\label{fig:2Dfig}
\end{figure}

Since the boundary edge between two adjacent triangles has the same length in each triangle, and the geometry internal to each triangle is Euclidean, the geometry from one triangle can be smoothly developed across an edge to any adjacent triangle. Intrinsic deviations from Euclidean 2-space only occur when a closed path contains one or more vertices. The intrinsic curvature is then constructed from the deficit angle $\epsilon$ around a vertex, given by the difference between the sum of the angles at the vertex and a full rotation in Euclidean space, see figure \ref{fig:2Dfig}.

\begin{figure}[h]
\begin{center}
\subfigure[Smooth Integrated Curvature]{
\includegraphics[trim = 10mm 0mm 10mm 0mm, clip, scale=0.5]{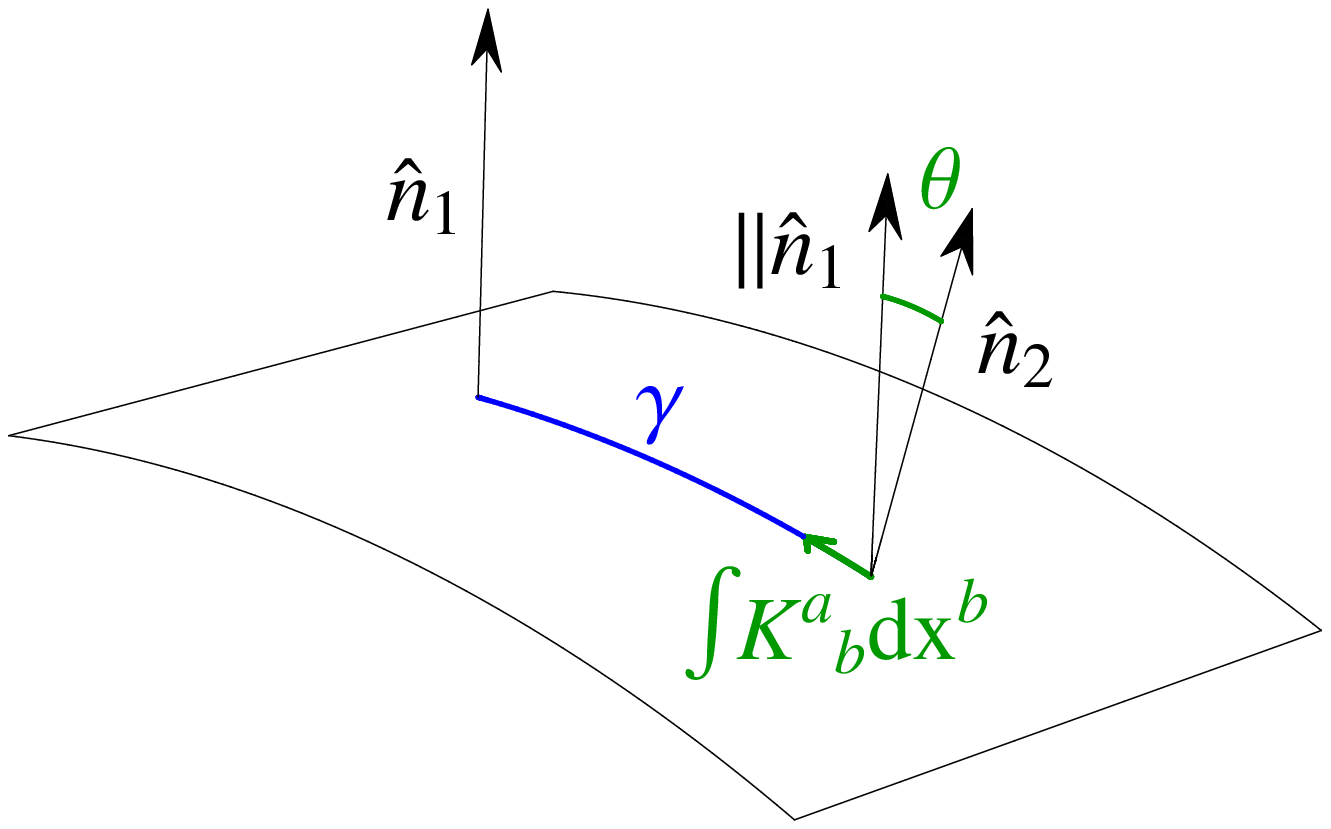} 
\label{fig:IntKab}
}
\subfigure[Simplicial Integrated Curvature]{
\includegraphics[trim = 20mm 30mm 10mm 0mm, clip, scale=0.5]{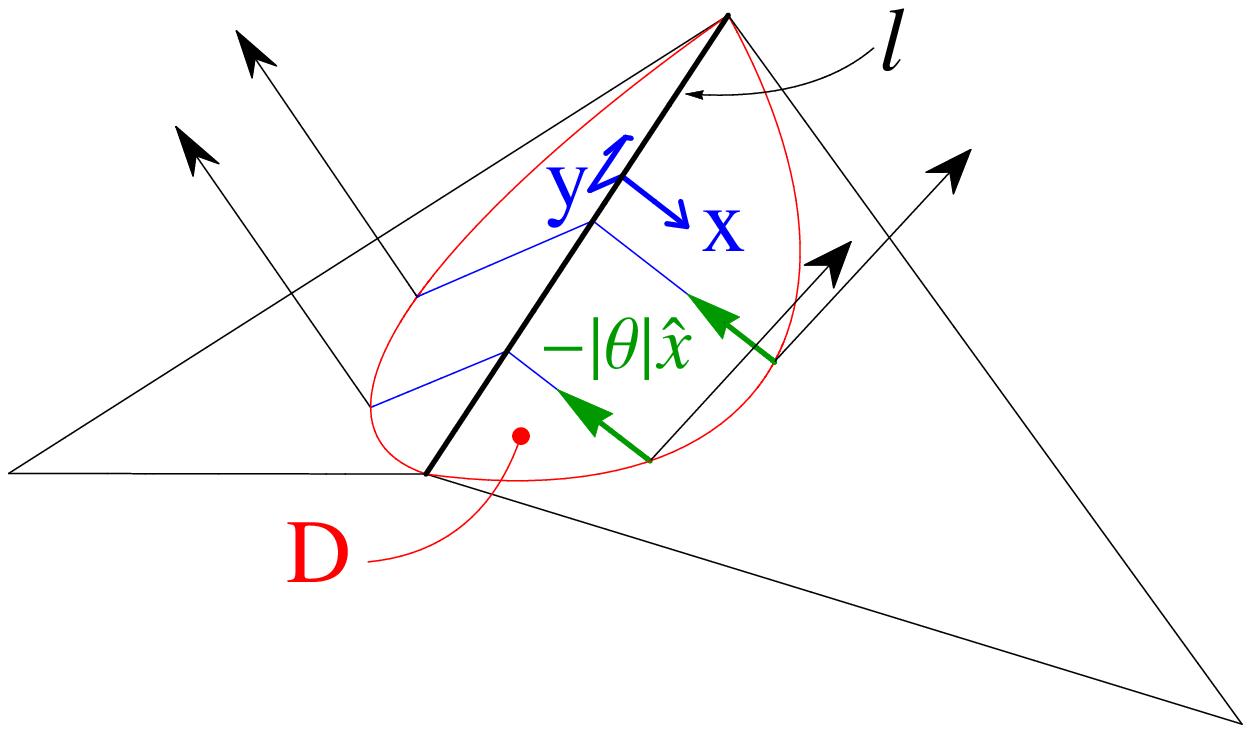} 
\label{fig:SimpIntK}
}
\end{center}
\caption{Integrated curvature along a finite path $\gamma$ on a smooth surface (a), and across an edge $l$ of a simplicial surface (b).}
\end{figure}

On a smooth surface in $\mathbb{E}^3$, the extrinsic curvature $K^a_b$ can be defined as the infinitesimal parallel transport of the unit normal $\hat n$, projected onto the surface. The sign convention of \emph{Misner, Thorne \& Wheeler} \cite{MTW} will be used here.
As shown in figure \ref{fig:IntKab}, integrating the curvature along a path on the surface gives
the \emph{finite} parallel transport of $\hat n_1$, projected onto the surface at the point $x_2$:
\begin{equation}
\int_{x_1}^{x_2} K^a_b \, d x^b \
  = \ \sin \theta \, \hat v^a \
  \simeq \ \theta \, \hat v^a \ .
\end{equation}
The unit vector $\hat v^a$ lies on the surface, in the direction of the projection of the vector $\hat n_1 - \hat n_2$
at the point $x_2$. Notice that if the surface in figure \ref{fig:IntKab} is concavely curved, the vector $\hat v^a$ will point in the opposite direction. The small angle approximation for $\sin \theta$ can be taken whenever the curvature changes slowly with respect to the distance between the two points.

On a piecewise-flat simplicial surface, the normal vector remains unchanged across a given triangle. The extrinsic curvature is thus concentrated on the edges $l$, or `hinges', between any pair of adjacent triangles. This is similar to the concentration of the \emph{intrinsic} curvature on the vertices. To find the extrinsic curvature associated with an edge $l$, an arbitrary convex region $D$ is defined, enclosing $l$ and entirely contained within the two triangles separated by $l$, see figure \ref{fig:SimpIntK}. Within $D$, an orthonormal coordinate basis $(x, y)$ is also defined, with $y$ parallel to the edge $l$. For a given value of $y$, the integral of $K^a_b$ along a path $\gamma(y)$,  parallel to the $x$ coordinate from one boundary to the other, is given by the difference $\theta$ between the normal vectors to each of the two triangles:
\begin{equation}
\int_{\gamma(y)} K^a_x \, dx \
  = \ \theta \, \hat v^a \
  = \ \theta \, \hat x \ .
\end{equation}
The variation between the two normals can only occur perpendicular to the `hinge' $l$, so the unit vector $\hat v^a$ must be parallel to the $x$-coordinate. The sign is then given by the orientation of the projection of $\hat n_1 - \hat n_2$
relative to the $x$-direction. This will be positive for a concave, and negative for a convex
hinging (as in figure \ref{fig:SimpIntK}). The small angle approximation here simply assumes that the lattice is dense enough to give a small variation in the normals between adjacent triangles.

Since the $y$ coordinate is parallel to $l$, integration of the extrinsic curvature along $y$ leaves the normal fixed, and so the $y$ components of $K^a_b$ can be assumed to vanish. The mean curvature $K$, given by the trace of the curvature tensor, is therefore equivalent to the $x x$ component. For a dense enough lattice, the integral of $K$ along $x$ is then given simply by the angle $\theta$ between the normal vectors of the two triangles, for any $y$:
\begin{equation}
\int_{\gamma(y)} K \, dx \ = \ \int_{\gamma(y)} K^x_x \, dx \ = \ \theta \ ,
\end{equation}
with $\theta$ again positive for a concave curvature, and negative for convex (as in the figures).
The integrated mean curvature over the entire region $D$ can now be found, since the integral of $K$ along the $x$ coordinate is invariant to the choice of $y$:
\begin{equation}
\int_D K \, d A \
 = \ \int_0^l \left[ \int_{\gamma(y)} K \ dx \right] d y \
 = \ \int_0^l \theta \, d y \
 = \ l \ \theta \ .
\end{equation}
Notice here that the label $l$ is used simultaneously to represent both a particular edge and the measure of that edge. This convention will be used for a number of labels, though it should be unambiguous which use is intended in each situation.

The value of the integrated mean curvature $l \, \theta$ does not depend on the shape of the region $D$, as long as it is convex, contains all of $l$, and does not cross any other edge. It is therefore possible to reduce the size of the region on either side of $l$, with the result remaining unchanged in the limit as $D$ converges to the edge $l$ itself. In this case, the integrated mean curvature associated with an edge can be defined as:
\begin{equation}\label{eq:IMCl}
IMC_l \ := \ l \, \theta \ ,
\end{equation}
depending simply on the length of the edge, and the angle between the normals of the two triangles on either side. This expression was derived from the boundary of the action by Hartle \& Sorkin \cite{HS81}, and by Brewin \cite{Brewin} using a similar argument to here, but employing a family of smooth surfaces limiting to the simplicial surface. In the context of Discrete Differential Geometry, this expression can also be found in \cite{DDGSul}, where it was credited to Jakob Steiner from 1840 \cite{Steiner}.

\section{Circumcentric Dual and Mean Curvature Density}
\label{sec:dual}

\begin{figure}[h]
\begin{center}
\subfigure[Circumcntric Dual Lattice]{
\includegraphics[scale=0.5]{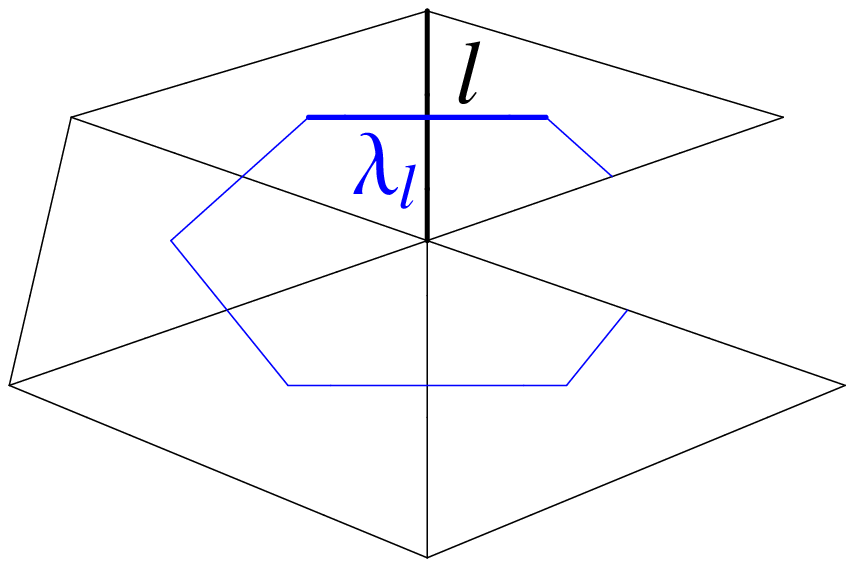} 
\label{fig:CircumDual}
}
\subfigure[Circumcentric Hybrid Area]{
\includegraphics[trim = 20mm 30mm 10mm 5mm, clip, scale=0.5]{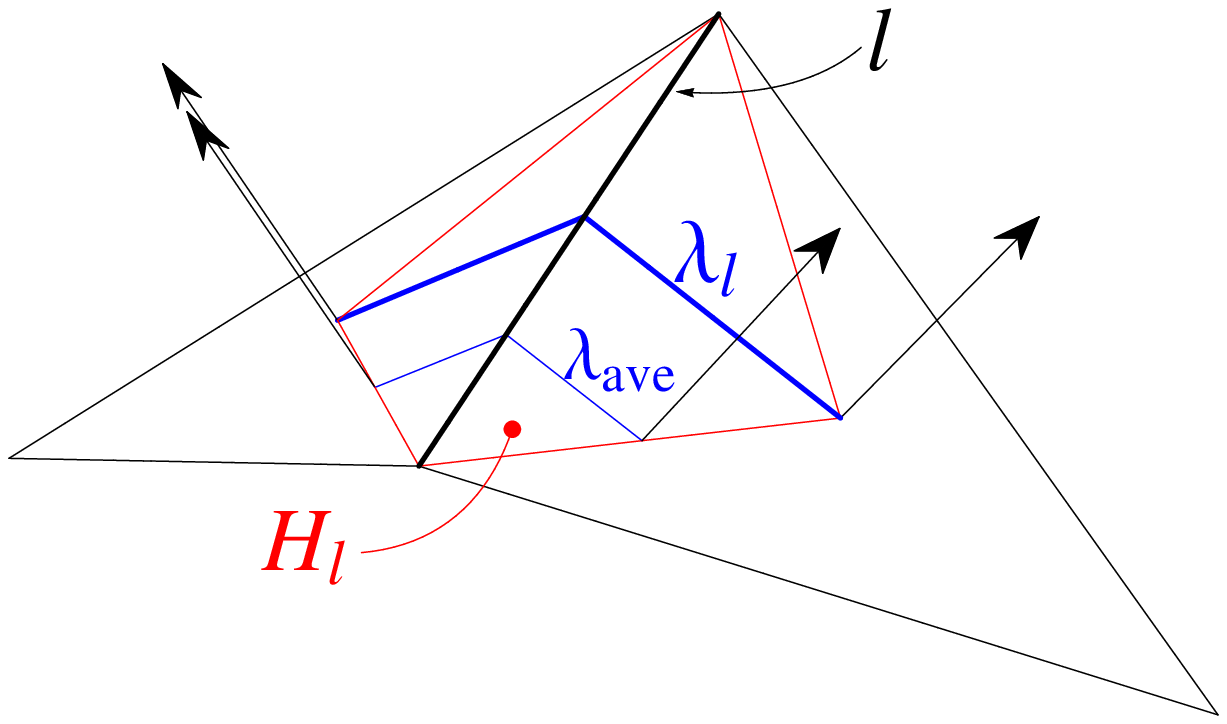} 
\label{fig:CircumHybrid}
}
\end{center}
\caption{The development of the simplicial surface from figure \ref{fig:2Dfig}, with the circumcentric dual edges shown (a), and the hybrid cell formed by the edge $l$ and its circumcentric dual $\lambda_l$ (b).}
\label{fig:Circum}
\end{figure}

To find a distribution of the curvature tensor over the entire surface, the circumcentric dual lattice leads to a remarkably natural definition for the region $D_l$. This lattice consists of lines within the simplicial surface, joining the circumcenters of adjacent triangles, as illustrated in figure \ref{fig:Circum}. For two triangles separated by an edge $l$, the line joining the circumcenters of the two triangles can be shown to be perpendicular to $l$, intersecting it at its midpoint. This circumcentric line is considered \emph{dual} to the simplicial edge $l$, and denoted $\lambda_l$. Since the circumcenters of each triangle can be determined from the edge lengths of the triangle, the length of a dual edge $\lambda_l$ is determined entirely from the five edges of the two triangles it connects.

A hybrid cell $H_l$ is defined as the quadrilateral with diagonals given by $l$ and $\lambda_l$. These cells form a complete tessellation of the simplicial surface. The internal geometry of each hybrid cell is Euclidean, formed by the development from one of the triangles intersecting the cell, across the edge $l$. One of the biggest advantages of the circumcentric hybrid cells, comes from the orthogonality of $l$ and $\lambda_l$. Unit vectors in these directions form a natural orthonormal basis for $H_l$, defining a piecewise-linear Cartan frame on $S$. The area of a hybrid cell is also conveniently given by the equation:
\begin{equation}\label{eq:areaHl}
  H_l \ = \ \frac{1}{2} \, l \, \lambda_l \ ,
\end{equation}
due to the orthogonality of $l$ with its circumcentric dual $\lambda_l$.

Defining the arbitrary region $D_l$ by the circumcentric hybrid volume $H_l$ leads naturally to a \emph{distributed} mean curvature, spread evenly over $H_l$:
\begin{equation}\label{eq:Kl}
K_l \
  = \ \frac{IMC_l}{H_l} \
  = \ \frac{l \, \theta}{\frac{1}{2} \, l \, \lambda_l} \
  = \ 2 \, \frac{\theta}{\lambda_l} \ .
\end{equation}
The average distance from one side of $H_l$ to the other, orthogonal to $l$, is given by half the length of the dual edge $\lambda_l$. This gives an alternative view of this distributed curvature, as the change in the normal vector from one side of $H_l$ to the other, divided by the average distance across the cell:
\begin{equation}\label{eq:Klave}
K_l \ = \ \frac{\theta}{d_{ave}} \
  = \ \frac{\theta}{\frac{1}{2} \lambda_l} \
  = \ \frac{2 \, \theta}{\lambda_l} \ .
\end{equation}
This gives an analogue of the continuous notion of the curvature as the rate of change of the normal vector with respect to a displacement along the surface. The role played by this average distance is again a special case of the circumcentric hybrid cells, resulting from its linear boundaries, and $\lambda_l$ as the maximum extension of $H_l$ orthogonal to the edge $l$.

The mean curvature associated with a triangle itself can now be found. A first requirement is that the integrated mean curvature over \emph{all} of the triangles of $S$, is equivalent to the sum of the integrated mean curvatures associated to each of the edges:
\begin{equation}\label{eq:IMC 2D}
IMC \ = \ \sum_t K_t \, A_t \
  = \ \sum_l IMC_l \ ,
\end{equation}
with the subscripts $t$ representing each triangle in the simplicial manifold. Using the densitised mean curvature given by the circumcentric hybrid cells:
\begin{eqnarray}\label{eq:IMC sum t 2D}
\sum_t K_t \, A_t \
  &= \ \sum_l IMC_l \nonumber \\
  &= \ \sum_l K_l \, H_l \nonumber \\
  &= \ \sum_t \left(\sum_{l \in \partial t} K_l \, H_{l | t} \right) \ ,
\end{eqnarray}
where $\partial t$ represents the boundary of the triangle $t$, and $H_{l | t}$ represents the restriction of the hybrid cell $H_l$ to that part which lies in the triangle $t$. These new cells $H_{l | t}$ are known as \emph{reduced} hybrid cells. The sum over all triangles $t$ can now be dropped from both sides, with the integrated mean curvature for a given triangle given by:
\begin{equation}\label{eq:IMCt}
IMC_t \ = \ K_t \, A_t \
  = \ \sum_{l \in \partial t} K_l \, H_{l | t} \ ,
\end{equation}
seen as an average of the distributed curvatures $K_l$ for the hybrid cells intersecting the triangle $t$, weighted according the areas $H_{l|t}$ of the restriction of their associated hybrid cells to $t$. With an expression for $K_l$ from \eref{eq:Kl} and \eref{eq:Klave}, and since $H_{l|t}$ is simply a triangle itself:
\begin{equation}\label{eq:IMCt ave lambda}
IMC_t \
  = \ \sum_{l \in \partial t} K_l \, H_{l | t} \
  = \ \sum_{l \in \partial t}
  \frac{2 \theta_l}{\lambda_l} \, \frac{l \, \lambda_{l | t}}{2} \
  = \ \sum_{l \in \partial t} l \, \theta_l \, \frac{\lambda_{l | t}}{\lambda_l} \ ,
\end{equation}
which is equivalent to the sum of the integrated mean curvatures at each of the edges, now weighted simply by the fraction of their associated circumcentric duals $\lambda_l$ that lie within the triangle $t$. A distributed mean curvature can also be defined for each triangle $t$, by spreading the integrated mean curvature evenly over the area of the triangle.

As mentioned earlier, the edge lengths for a simplicial manifold cannot be chosen arbitrarily, with the triangle inequality requiring satisfaction. For the circumcentric dual to give a Voronoi lattice, i.e. to subdivide $S$ into regions of points closer to a single vertex than to any other, the triangles must also satisfy the Delaunay condition. For situations where this is not true, see \cite{HambWill} and Appendix A of \cite{DyerPhD}. However for a surface embedded in $\mathbb{E}^3$, there are procedures for forming Delaunay triangulations from a given triangulation \cite{Tess,BDGintrins}.

The intuition used above also works best for `well-centered' triangulations, where the circumcenter of each triangle is contained within the triangle. This is stronger than the Delaunay condition. A triangle which is \emph{not} well-centered will have a ristricted dual edge which lies outside the triangle. For a consistent integrated mean curvature, this restricted edge must be considered to have a \emph{negative} length. This leads to a reduced hybrid cell, again lying outside the triangle, with negative area. However, this should not cause much difficulty for reasonably smooth triangulations, where there is no large difference between the curvatures at different edges of each triangle.

\section{Higher Dimensions and Non-Euclidean Manifolds}
\label{sec:HigherD}

In this section it is shown that there is no essential difficulty in extending the circumcentric based distribution of the extrinsic curvature in higher dimensions, and more general embeddings. The generalization of a piecewise-flat simplicial manifold of arbitrary dimension $S^n$, without reference to any embedding, is first described in section \ref{sec:HigherDS}. This includes higher dimensional dual lattices and hybrid cells. The distribution of the extrinsic curvature for an embedding in $\mathbb{E}^{n+1}$ is given in section \ref{sec:hyper}, requiring very little adjustment from the $2$-dimensional case. More general situations are then discussed briefly in section \ref{sec:HDgeneral}.

\subsection{Higher dimensional simplicial manifolds}
\label{sec:HigherDS}

In higher dimensions, the concept of a triangle is generalized to an $n$-simplex. In Euclidean space these are n-volumes, formed by $n + 1$ vertices, with straight lines joining each pair of vertices. In one, two and three dimensions, $n$-simplices are given by line segments, triangles and tetrahedra respectively. The boundary of each $n$-simplex is formed by a collection of $(n - 1)$-simplices, $n + 1$ in number. For example a triangle is bounded by $3$ line segments, and a tetrahedron by $4$ triangles. These $(n - 1)$-simplices must also satisfy a generalization of the triangle inequality, to ensure that the simplex is well defined, and that its $n$-volume is non-zero. The great advantage of $n$-simplices is that their geometry in Euclidean space is entirely determined by the lengths of their edges.

An $n$-dimensional piecewise-flat simplicial manifold $S^n$ is a manifold which can be decomposed into a collection of flat n-simplices, connected to each other by their boundary $(n - 1)$-simplices. These are the $n$-dimensional equivalent of approximating a 1-dimensional curve by straight line segments. Since each $n$-simplex is determined entirely by its edge lengths, the internal geometry of $S^n$ is again determined completely by the lengths of all of its edges $l$.

The Euclidean geometry interior to each $n$-simplex can again be developed across its $(n - 1)$-simplices to any adjacent $n$-simplex. The deviation of the internal geometry of $S^n$ from Euclidean space now shows up in the deficit angles around an $(n - 2)$-simplex. The circumcentric dual lattice is formed by edges connecting the circumcenters of the $n$-simplices, passing perpendicularly through the circumcenters of the $(n - 1)$-simplices. These form the `hinges' $h$ on which the \emph{extrinsic} curvature of $S^n$ is concentrated. The corresponding circumcentric dual edge to an $(n - 1)$-simplex $h$ is denoted $\lambda_h$.

\begin{figure}[h]
\begin{center}
\subfigure[Two tetrahedra and their hybrid cell]{
\includegraphics[scale=0.4]{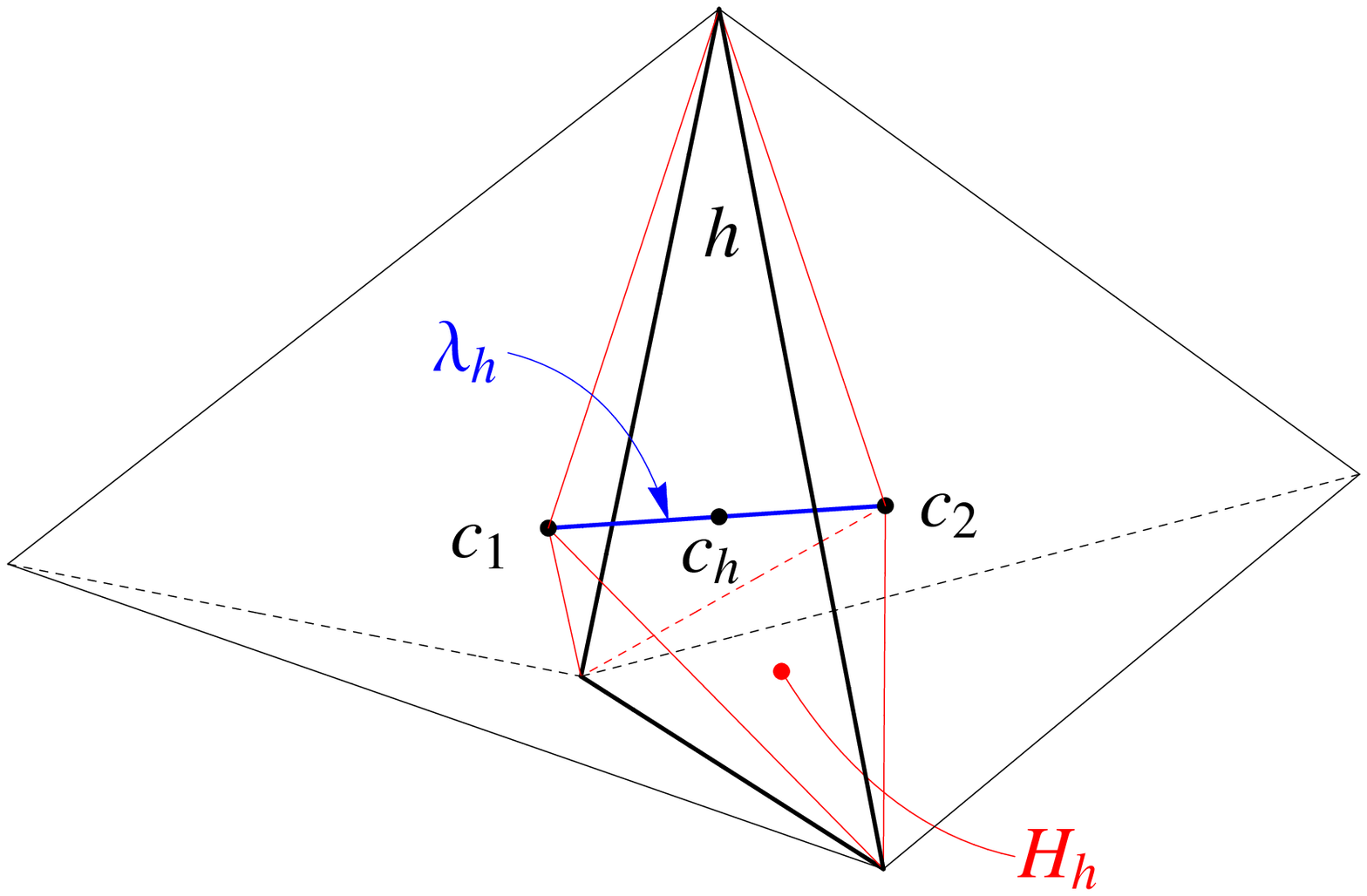} 
\label{fig:3SimplexA}
}
\subfigure[Hybrid cell ($\lambda_h$ direction expanded)]{
\includegraphics[scale=0.4]{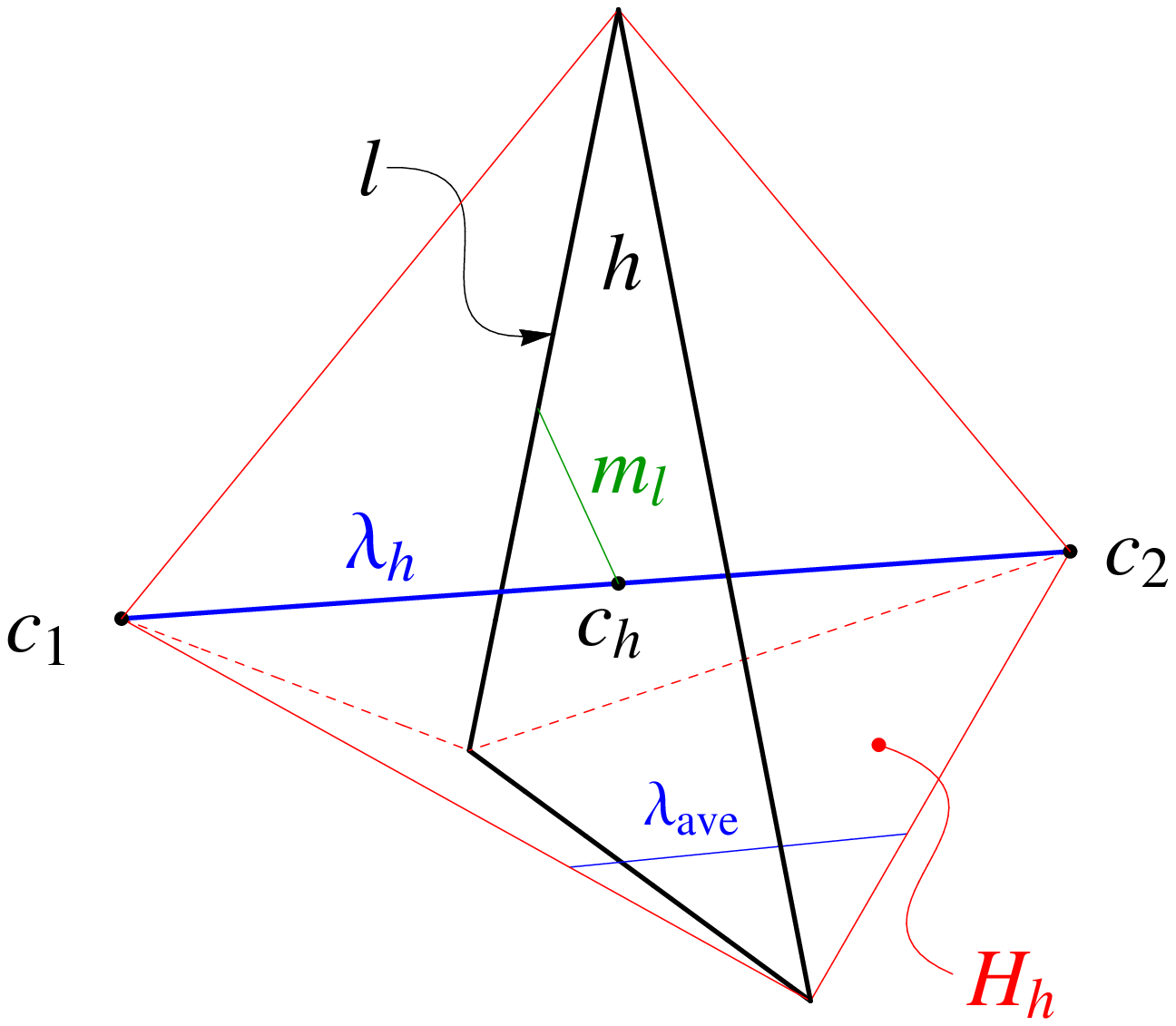}
\label{fig:3SimplexB}
}
\end{center}
\caption{Two tetrahedra, with the hinge $h$ given by the triangle separating them, and the dual edge $\lambda_h$ joining the circumcenters of the two tetrahedra $c_1$ and $c_2$, (a). The dual edge can also be seen to pass through the circumcenter $c_h$ of $h$. In (b), the average distance $\lambda_{ave}$ across $H_h$ perpendicular to $h$ is displayed. The line $m_l$ joining the circumcenter of the hinge $c_h$ to the center of the edge $l$ is also shown. This line is perpendicular to both $\lambda_h$ and $l$. The triplet $(l, m_l, \lambda_h)$ thus form an orthonormal frame within $H_h$.}
\label{fig:3Simplex}
\end{figure}

The $n$-dimensional hybrid cell $H_h$ is formed by joining the vertices of $h$ with those of $\lambda_h$, forming a complete tessellation of $S^n$. An orthonormal basis can again be set up within each $H_h$, consisting of a unit vector along $\lambda_h$, a unit vector along the line joining the circumcenter of $h$ to the circumcenter of one of the $(n - 2)$-simplices bounding it, and continuing until an edge $l$ is reached. This can be seen in figure \ref{fig:3Simplex} for the 3-dimensional case, also see \cite{SRF} for more details. This again defines a piecewise-linear Cartan frame, or $n$-bein on $S^n$, with some degeneracy in the choice of unit vectors interior to $h$ (in figure \ref{fig:3Simplex} there are three choices for the pair $m_l$ and $l$). Due to the orthogonality of $h$ and $\lambda_h$, the volume of $H_h$ can easily be computed:
\begin{equation}\label{eq:volumeHh}
H_h \ = \ \frac{1}{n} \, h \, \lambda_h \ .
\end{equation}
The simplicity of this computation is a major advantage of these hybrid cells.

Care should again be taken to ensure that $S^n$ gives a Delaunay triangulation, although this gets more difficult with increasing dimension. Recent progress has been made in perturbing given triangulations to form Delaunay triangulations on non-Euclidean manifolds however, irrespective of any embedding \cite{BDGDelaun,DVW}. Again, the intuition in the following relies on circumcenters lying in the interior of $n$-simplices, however reasonable deviations should not cause problems.

\subsection{Simplicial hypermanifolds in $\mathbb{E}^{n + 1}$}
\label{sec:hyper}

For an $n$-dimensional simplicial manifold $S^n$, embedded in a Euclidean space of one extra dimension $\mathbb{E}^{n + 1}$, there is a unique normal vector to each $n$-simplex. The arguments for the 2-dimensional case apply in the same way here. The angle $\theta$ between the normal vectors on either side of an $(n - 1)$-simplex $h$, occur only in the direction orthogonal to $h$, i.e. in the direction of the dual edge $\lambda_h$. The integral of the mean curvature $K$ from one side of $h$ to the other, is given by the integral of the $K^\lambda_\lambda$ component of the curvature, and equal to $\theta$. Defining a convex region $D$, enclosing $h$ and no other $(n-1)$-simplex, the integrated mean curvature over $D$ is:
\begin{equation}
\int_D K \, d V^n \
 = \ \int_h \left[ \int_{\gamma} K \ d\lambda \right] d V^{n-1} \
 = \ \int_h \theta \ d V^{n-1} \
 = \ h \ \theta \ ,
\end{equation}
with the inner integral in the second part taken along each path $\gamma$ parallel to $\lambda_h$.

As in the $2$-dimensional case, this equation is unchanged in the limit as $D$ converges on $h$, so that the integrated mean curvature along a hinge can be taken as:
\begin{equation}\label{eq:IMCh}
IMC_h \ = \ h \, \theta \ ,
\end{equation}
generalizing \eref{eq:IMCl} to higher dimensions. Instead, however, taking the region $D$ as the circumcentric hybrid cell $H_h$ seems to give a much richer interpretation. A distributed mean curvature is given by spreading $IMC_h$ evenly over $H_h$:
\begin{equation}\label{eq:Kh}
K_h \ = \ \frac{IMC_h}{H_h} \
 = \ \frac{h \, \theta}{\frac{1}{n} \, h \, \lambda_h} \
 = \ \frac{n \, \theta}{\lambda_h} \ ,
\end{equation}
giving a more general form of \eref{eq:Kl}. This can again be viewed as the integrated mean curvature along the \emph{average} distance across $H_h$, orthogonal to $h$. Integrating the distances along a $\hat \lambda_h$ direction within $H_h$, over the $(n-1)$-simplex $h$, gives the volume of the hybrid cell $H_h$ itself. The average distance $d_{ave}$ is then found by dividing again by the volume of $h$, giving a densitised mean curvature of:
\begin{equation}\label{eq:Khave}
K_h \ = \ \frac{\theta}{d_{ave}} \
 = \ \frac{\theta}{H_h / h} \
 = \ \frac{\theta}{\frac{1}{n} \, \lambda_h} \
 = \ \frac{n \, \theta}{\lambda_h} \ ,
\end{equation}
using \eref{eq:volumeHh} for the volume of $H_h$. This is equivalent to \eref{eq:Kh} above, and generalizes \eref{eq:Klave}, giving a discrete analogue of the infinitesimal extrinsic curvature.

The integrated curvature over an $n$-simplex $s$ can be found using \eref{eq:IMC 2D}, \eref{eq:IMC sum t 2D} and \eref{eq:IMCt}, with the appropriate $n$-dimensional substitutions, giving:
\begin{equation}\label{eq:IMCs}
IMC_s \ = \ \sum_{h \in \partial s} K_h \, H_{h|s} \ ,
\end{equation}
with the reduced hybrid cell $H_{h|s}$ representing the restriction of $H_h$ to that part which lies in $s$. The result of \eref{eq:IMCt ave lambda} also generalizes, from \eref{eq:volumeHh} and \eref{eq:Kh}:
\begin{equation}\label{eq:IMCs ave lambda}
IMC_s \ = \ \sum_{h \in \partial s} h \, \theta_h \ \frac{\lambda_{h|s}}{\lambda_h} \
 = \ \sum_{h \in \partial s} \frac{\lambda_{h|s}}{\lambda_h} \ IMC_h \ ,
\end{equation}
with the integrated mean curvature over an $n$-simplex $s$ given by the sum of the integrated mean curvatures associated with each of its bounding $(n-1)$-simplices $h$, from \eref{eq:IMCh}, weighted according to the fraction of the corresponding dual edges $\lambda_h$ which lie in $s$.

\subsection{More general submanifolds}
\label{sec:HDgeneral}

The definitions can be generalized for submanifolds of non-Euclidean manifolds $M$ in the usual way. The simplicial manifold $S^n$ may be considered as an approximation of a smooth submanifold $H$ of $M$, with the edge lengths in $S^n$ defined by geodesic lengths in $H$, and parallel transport of normal vectors requiring a connection on $M$.

The simplicial manifold $S^n$ may also be embedded in a larger piecewise-linear simplicial manifold $S^{n+m}$. In this case each simplex in $S^n$ lies on the boundary of some $(n+1)$-simplex in $S^{n+m}$. The results of section \ref{sec:hyper} can be directly applied here, due to the flat space on the interior of each $(n+1)$-simplex, and its development around a hinge of $S^n$.

For submanifolds with a higher codimension, for each $n$-simplex $s$ there is a unique normal vector $\hat n_s$ from which the extrinsic curvature can be defined. The complete embedding of $S^n$ will also require information about the change of $span \{T_s S, \hat n_s\}$ across a hinge $h$. However this is no more complicated than a $1$-dimensional curve in $\mathbb{E}^3$, as described using the Seret-Frenet frame, see \cite{DDGSul} for example.


\section{Conclusion}

The circumcentric hybrid cells $H_h$ have been shown to give a very natural volume for distributing the local integrated mean curvature. This gives a pointwise mean curvature for each hybrid cell, which completely determins the extrinsic curvature within each cell. These circumcentric hybrid cells have a number of clear advantages over other volumes: $(i)$ they provide a complete tessellation of the simplicial manifold, $(ii)$ a geometric orthonormal basis is provided by their circumcentric construction, giving a particularly straight-forward expression for the volume \eref{eq:volumeHh}, and $(iii)$ the average distance parallel to the dual edge gives the precise distance for a fractional rate of change of the normal vectors \eref{eq:Khave}.

The use of circumcentric hybrid cells for the distributed mean curvature fits well with a broader use of the circumcentric dual lattice in Regge calculus. The circumcentric polytopes give a natural area for distributing the Gaussian curvature, and the other circumcentric hybrid cells have been shown to give consistent definitions for the Riemann and Ricci tensor, and the Riemann scalar curvatures \cite{MMR,SRT}. The circumcentric dual lattice arrises naturally in the Null Strut Calculus, where light rays (null struts) from the vertices of each spacelike tetrahedron meet along the timelike normal based at the circumcenter of the tetrahedron \cite{KLM89ns}. It is also consistent with the use of circumcentric dual lattices in Discrete Differential Geometry, and particularly in Discrete Exterior Calculus \cite{HiraniPhD,MMYec}.

\ack

We acknowledge support for this research from Air Force Research Laboratory Grant $\#$ FA8750-15-2-0047.

\section*{References}

\bibliography{SimpGeom}
\bibliographystyle{unsrt}

\end{document}